\documentclass[prl,floatfix,twocolumn,showpacs,preprintnumbers,amsmath,amssymb,superscriptaddress]{revtex4}
\usepackage{graphicx,psfrag} 
\begin{document}   
\title{Improved Higgs Naturalness With or Without Supersymmetry} 
\date{October 12, 2006}
\author{Ben Gripaios}  
\email{b.gripaios1@physics.ox.ac.uk}  
\affiliation{Rudolf Peierls Centre for Theoretical Physics, University of Oxford,
1 Keble Rd., Oxford OX1 3NP, UK}
\affiliation{Merton College, Oxford OX1 4JD, UK}
\affiliation{Temporary address: Dept.\ of Physics, University of California, Davis, CA 95616, USA}
\author{Stephen M. West}
\email{s.west1@physics.ox.ac.uk}
\affiliation{Rudolf Peierls Centre for Theoretical Physics, University of Oxford,
1 Keble Rd., Oxford OX1 3NP, UK}
\affiliation{Temporary address: Dept.\ of Physics, University of California, Davis, CA 95616, USA}
\begin{abstract}
Many models of electroweak symmetry-breaking with an extended Higgs sector exhibit improved naturalness, wherein the new physics scale, at which quadratic divergences of Higgs mass parameters due to top quark loops are cut off, can be pushed beyond the reach of the Large Hadron Collider without unnatural fine tuning. Such models include examples where the new physics is supersymmetry, implying that supersymmetry may still solve the hierarchy problem, even if it eludes detection at the LHC.
\end{abstract}   
\preprint{OUTP-0607P}
\maketitle 
\section{\label{const}Constraints on the Scale of New Physics}
The standard model (SM) of particle physics provides an excellent fit to precision electroweak experiments, provided the mass $m_H$ of the as-yet unobserved Higgs boson is rather light \cite{unknown:2005em},
\begin{gather} \label{ewsm}
m_H \lesssim 285\  \text{GeV}\  (95\ \% \  \text{C.\ L.\ }).
\end{gather}
However, the SM suffers from the hierarchy problem: the electroweak scale is sensitive to higher energy scales occurring in nature, via a quadratically-divergent contribution to the Higgs mass parameter coming dominantly from virtual top quarks.  The divergence must be cut off by new physics at some energy scale $\Lambda_t$, which, in the absence of unnatural fine-tuning of parameters, should also be rather low.
This has been a major motivation for the ongoing construction of the Large Hadron Collider (LHC). A quantitative measure of the fine-tuning constraint may be given as \cite{Ellis:1986yg}
\begin{gather} \label{ftsm}
\Lambda_t \lesssim  400\ \text{GeV} \left( \frac{m_H}{115\ \text{GeV}} \right) \sqrt{D_H},
\end{gather}
where $D_H = \partial \log m_H^2 / \partial \log \Lambda_t^2$ and the amount of fine-tuning is roughly one part in $D_H$; in a natural theory, $D_H$ should be of order unity and the new physics will thus be accessible at the LHC.

The minimal supersymmetric extension of the standard model (MSSM), in which the quadratic divergence is elegantly cut off by the superpartners of the top quarks (the stops), is already unnatural in this sense: the stops must have masses of many hundreds of GeV in order to push the mass of the lightest Higgs boson above the empirical lower bound obtained from direct searches at LEP \cite{Inoue:1982pi} and the fine-tuning is at least a few {\it per cent} \cite{Luty:2005sn}.

However, the electroweak (\ref{ewsm}) and fine-tuning (\ref{ftsm}) constraints are changed in a theory with an extended Higgs sector. 
Recently, Barbieri and Hall \cite{Barbieri:2005kf} showed that even the simplest extension of the SM, the (non-supersymmetric) two Higgs doublet model, exhibits regions of parameter space with `improved naturalness', in which the combined electroweak and fine-tuning constraints are relaxed, allowing the scale of new physics to be as large as 2 TeV or more. This leads to the rather pessimistic conclusion that the new physics, whatever it may be, could lie beyond the reach of the LHC. 

In the following, we first wish to point out that the phenomenon of improved naturalness is rather generic, in that many such models should exist. We then exhibit a model of this type in which the new physics is supersymmetry. Again, many such models should exist. We conclude, therefore, that even if no  direct evidence for supersymmetry is seen at the LHC, it may still provide the solution to the hierarchy problem \cite{note1}.

\section{\label{many}Improved Naturalness from Extended Higgs Sectors}
The only apparent prerequisite for improved naturalness is an extended Higgs sector, with multiple Higgs doublets and possibly singlets. (We exclude higher-dimensional representations of $SU(2)_L$, which generically violate the electroweak constraints if they acquire vacuum expectation values [VEVs]). 
In such a model, the Higgs mass eigenstates that couple significantly to gauge bosons (and are thus constrained by electroweak precision tests) need not coincide with the eigenstates which are sensitive to $\Lambda_t$. Indeed, the overlap between the different eigenstates (and {\it ergo} the constraints) is necessarily complete only in the theory with a single Higgs boson, {\it viz.} the SM. 

What is more, the individual constraints generalizing (\ref{ewsm}) and (\ref{ftsm}) are themselves changed in an extended Higgs sector and may also relax the constraint on the scale of new physics.

Now, starting from the two Higgs doublet example of Barbieri and Hall, we see that there should exist many models with improved naturalness and a scale of new physics beyond the reach of the LHC \cite{note2}.
Indeed, consider all models that reduce at low energies to the example of Barbieri and Hall in some decoupling limit in which
one or more Higgs bosons become very massive and decouple from gauge bosons.  In this limit, the heavy Higgs states decouple from electroweak physics by {\it fiat}  and therefore do not affect the electroweak constraints. Nor are their masses finely-tuned. The upper bound on the natural scale of new physics in such a model can thus be pushed at least as high as in \cite{Barbieri:2005kf}, by going to the decoupling limit, if not higher. 

Large though this class of models which automatically incorporate improved naturalness may be, it is not obvious that it includes models in which the new physics is supersymmetry. But the existence of such a model would be of paramount importance, implying that low-energy supersymmetry as the solution of the hierarchy problem could not be ruled out, even if no direct evidence for it were to be observed at the LHC. We now present such a model, the so-called `fat Higgs' model of \cite{Harnik:2003rs}. 

\section{A Supersymmetric Model with Improved Naturalness}
Models with supersymmetry necessarily have multiple Higgs doublets and are, as such, candidates for having improved naturalness. 
A supersymmetric theory containing only multiple Higgs doublet superfields (such as the MSSM) is hampered, though, by the fact that the quartic couplings in the scalar potential for Higgs bosons are fixed, at tree-level, by the gauge couplings and the lightest neutral Higgs boson mass is consequently rather light \cite{Inoue:1982pi,Drees:1988fc}. In order to push the mass above the bounds set by LEP, one is forced into a region of parameter space where the lightest mass is strongly sensitive to the scale of new physics (which is given by the stop masses in this case).

Some leeway is obtained by adding an electroweak-singlet chiral superfield to the model, which introduces an arbitrary quartic coupling $\lambda$ into the scalar potential \cite{gun1}. The standard lore \cite{Haber:1986gz,Drees:1988fc} is that this coupling, though a free parameter, should not be too large if it is to remain perturbative up to the gauge-coupling unification scale of $10^{16}$ GeV. However, successful unification of gauge couplings is not contingent upon $\lambda$ remaining perturbative \cite{Nom,Harnik:2003rs}: It is possible to allow the theory to go through a supersymmetric strong-coupling transition, whilst retaining gauge-coupling unification at higher energies (just as, for example, electroweak unification at around a hundred GeV
is not spoiled by the couplings of the low energy effective chiral Lagrangian for QCD becoming strongly-coupled at a GeV or so).
For the fat Higgs model \cite{Harnik:2003rs}, we expect $\lambda \simeq 4 \pi$ at the strong coupling scale $\Lambda$ on the basis of naive dimensional analysis \cite{Weinberg:1978kz}, while at a lower energy scale $\mu$, the coupling decreases according to the renormalization group equation 
\begin{gather} \label{RGE}
\lambda^2 (\mu) = \frac{16\pi^2}{1 + 8 \log \Lambda/\mu}.
\end{gather}

This has an important consequence: the upper bound on the lightest Higgs mass is lifted substantially (to $\lambda v \sin \beta$, where $v$ and $\beta$ are defined below), such that the region of parameter space of the model in \cite{Harnik:2003rs} allowed by direct searches at LEP is rather large. As we shall now see, it includes regions exhibiting improved naturalness, in which the masses of stops and other superpartners can be pushed beyond the reach of the LHC.  

Below the strong coupling scale, the model contains the Higgs doublet chiral superfields $H_u$ and $H_d$ of the MSSM together with a singlet superfield $N$. The superpotential contains the terms 
\begin{gather} \nonumber
 \lambda N (H_d H_u - v_0^2) 
\end{gather}
and the potential for the neutral Higgs scalars, $H_u^0$, $H_d^0$ and $N$, is \cite{Fayet}
\begin{gather} \label{pot}
V = V_F + V_D + V_{\text{soft}},
\end{gather}
where
\begin{align}
 V_{F} &= \lambda^2 |H_d^0 H_u^0 - v_0^2|^2 
  + \lambda^2 |N|^2 (|H_u^0|^2 + |H_d^0|^2),  \nonumber \\
V_D &=  \frac{g^2 + g^{\prime 2}}{8} (|H_u^0|^2 - |H_d^0|^2)^2 \quad \text{and}\nonumber \\
V_{\text{soft}} &= m_1^2 |H_d^0|^2 + m_2^2 |H_u^0|^2 + m_0^2 |N|^2 + \nonumber \\ 
&\phantom{=} + (A \lambda N H_d^0 H_u^0 - C \lambda v_0^2 N + \text{h.~c.}) \nonumber
\end{align}
represent contributions from supersymmetric $F$- and $D$-terms and soft supersymmetry-breaking terms, respectively. The parameters $v_0,A,C,m_0,m_1$ and $m_2$ have dimensions of mass.
Following \cite{Harnik:2003rs}, we take the quartic coupling $\lambda$ to be large compared to the gauge couplings $g$ and $g'$ and neglect the $D$-terms in what follows. To further simplify the analysis, we set $A=C=0$ \cite{note4}, as in \cite{Harnik:2003rs}. 

The neutral scalars acquire VEVs given by
\begin{align}
\langle H_u^0 \rangle &= v_0\sqrt{1-\frac{m_1 m_2}{\lambda^2v_0^2}}\sqrt{\frac{m_1}{m_2}} ,
  \nonumber \\
\langle  H_d^0 \rangle &= v_0\sqrt{1-\frac{m_1 m_2}{\lambda^2v_0^2}}\sqrt{\frac{m_2}{m_1}} \quad \text{and}
  \nonumber \\
\langle  N \rangle &= 0. \nonumber
\end{align}
It is convenient to define
\begin{gather}
m_s = \sqrt{m_1 m_2} \qquad \text{and} \qquad \tan\beta \equiv \frac{\langle H_u^0 \rangle}{\langle H_d^0 \rangle} 
  = \frac{m_1}{m_2}, \nonumber
\end{gather}
in terms of which the electroweak scale is given by
\begin{gather} \label{weak}
 v^2 \equiv {\langle H_u^0 \rangle}^2 + {\langle H_d^0 \rangle}^2 = 2 \frac{\lambda^2 v_0^2 - m_s^2}{\lambda^2 \sin 2\beta} = (174 \ \text{GeV})^2 .
\end{gather}

The Higgs boson mass spectrum is then as follows.
The masses of the charged Higgs scalars $H^\pm$, the singlet scalar and pseudoscalar $N$ and the doublet pseudoscalar $A$ are given by \cite{gun2}
\begin{align}
 m^2_{H^\pm} &= \frac{2 m_s^2}{\sin 2\beta}, \nonumber \\
m^2_{N} &= \lambda^2 v^2 + m_0^2 \quad \text{and} \nonumber \\
m^2_{A} &= \lambda^2 v^2 + m_{H^\pm}^2, \nonumber
\end{align}
respectively. The masses of the lighter and heavier doublet neutral Higgs scalars, $h^0$ and $H^0$ respectively, are given by the eigenvalues $m^2_{h,H}$ of the matrix
\begin{gather} \label{mm}
\begin{pmatrix}
\lambda^2 v^2 \cos^2 \beta + m_s^2 \cot \beta &  \lambda^2 v^2 \sin \beta \cos \beta - m_s^2 \\
 \lambda^2 v^2 \sin \beta \cos \beta - m_s^2 & \lambda^2 v^2 \sin^2 \beta + m_s^2 \tan \beta
\end{pmatrix}.
\end{gather}

We now discuss how the electroweak (\ref{ewsm}) and fine-tuning (\ref{ftsm}) constraints are modified in this model, beginning with the latter. The dominant sensitivity of weak-scale masses (namely, the $Z$-boson mass $m_Z$, $m_h$ and $m_H$) to the new physics scale comes from the mass parameter $m_2$, which receives a  one-loop correction from virtual tops and stops, given by \cite{Luty:2005sn}
\begin{gather} \label{loop}
\Delta m_2^2 = - \frac{3 y_t^2}{4 \pi^2} m_{\tilde{t}}^{2} \log \frac{\Lambda}{m_{\tilde{t}}},
\end{gather}
where $y_t$ is the top quark Yukawa coupling. The stop mass $m_{\tilde{t}}$ can then be written in terms of the weak-scale masses as
\begin{gather} \label{stop}
m^2_{\tilde{t}} \simeq \frac{8 \pi^2}{3}\frac{ v^2\sin^2 \beta}{ m^2_t} \frac{m^2_{h,H,Z} }{\partial m_{h,H,Z}^2 / \partial m_2^2 } \frac{D_{h,H,Z}}{\left(2 \log \Lambda /m_{\tilde{t}}-1\right)},
\end{gather}
where $m_t$ is the mass of the top quark.
The derivatives $\partial m_{Z}^2 / \partial m_2^2$ and $\partial m_{h,H}^2 / \partial m_2^2$ can be determined from (\ref{weak}) and (\ref{mm}), and are $O(\lambda^{-2})$ and $O(1)$, respectively.

The three fine-tuning parameters $D_{h,H,Z} = \partial \log m_{h,H,Z}^2/\partial \log m_{\tilde{t}}^2$ should not be much larger than unity in a natural theory \cite{Ellis:1986yg}, but a large stop mass can nevertheless be obtained as follows. Firstly, the Higgs masses $m_{h,H}$ can be large, and the $Z$-boson mass, though fixed at the weak scale, is accompanied by an extra factor of $\lambda$, coming from the derivative in the denominator of (\ref{stop}).
Secondly, the logarithm in the denominator, which is given by evaluating (\ref{RGE}) at $\mu=m_{\tilde{t}}$, is not large. (For the range of $\lambda$ that we consider below, $5 < \Lambda / m_{\tilde{t}} < 10$, giving a reasonable separation between the supersymmetric strong-coupling scale and supersymmetry-breaking masses.) These two effects conspire to lift the natural stop mass well above the value that would be obtained in a supersymmetric theory which remains perturbative all the way up to the unification scale, in which $m_h$ is light and the logarithm in (\ref{loop}) is large enough to cancel the loop factor.

To what extent is the improved naturalness region consistent with constraints from electroweak precision tests, generalizing (\ref{ewsm})? The Higgs scalars have masses over several hundred GeV and superpartners have masses of several TeV. We therefore expect the oblique approximation, including just $S$ and $T$ parameters \cite{Peskin:1991sw}, to be a good one \cite{Wells:2005vk}. 
(As exceptions, we do include the constraints from vertex corrections to $\Gamma (Z \rightarrow b\bar{b})$ and $b \rightarrow s\gamma$ \cite{Haber:1999zh}.)
Moreover, since all superpartners are very massive, they do not contribute significantly to $S$ and $T$ \cite{Martin:2004id} and we need only include the contributions from Higgs scalars. 

\begin{figure}
\includegraphics[width=7.5cm]{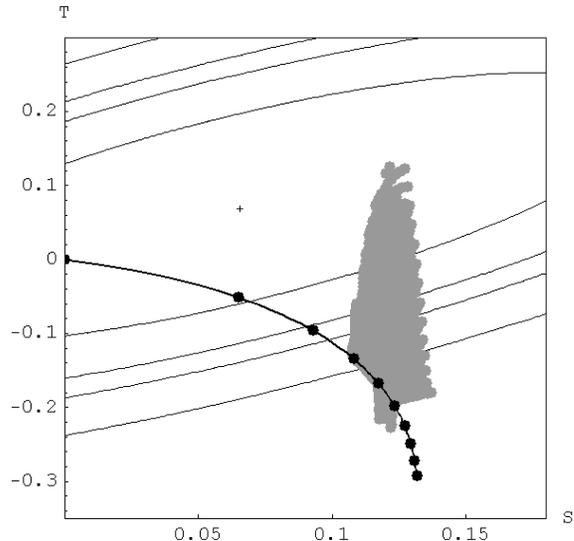}
\caption{\label{st}Constraints on the $S$ and $T$ parameters from electroweak data. The elliptical contours, centered on the best-fit values (+), enclose the 68, 90, 95 and 99 \% confidence regions. The grey area shows the model contribution for points in parameter space with $m_{\tilde{t}}> 2$ TeV and satisfying the $\Gamma (Z \rightarrow b\bar{b})$ and $b \rightarrow s\gamma$ constraints. For comparison, the thick line indicates the SM contribution with Higgs masses between 100 GeV (at the origin) and 1 TeV in increments of 100 GeV.}
\end{figure}
\begin{figure}
\includegraphics[width=7.5cm]{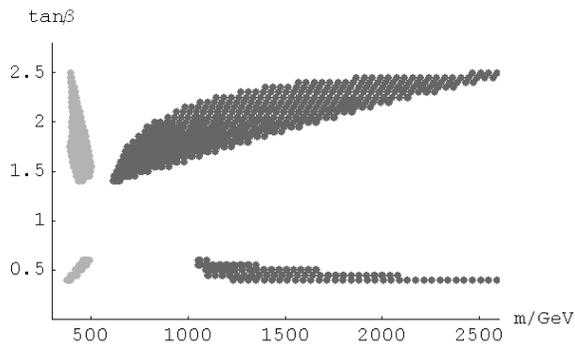}
\caption{\label{hb}Masses of the light (heavy) Higgs scalars {\it vs.} $\tan \beta$, where the light (dark) shaded regions have $m_{\tilde{t}}> 2$ TeV and satisfy precision electroweak tests at the 95\% C.L..}
\end{figure}

To fit the electroweak data, we follow the procedure of \cite{Harnik:2003rs} verbatim, fitting $S$ and $T$ to the three best-measured observables, {\it viz.} $m_W$ (the mass of the $W$), $\Gamma_l$ ( the leptonic width of the $Z$) and $\sin^2 \theta^{\text{lept}}_{\text{eff}}$ (the weak mixing angle appearing in $Z$ decay asymmetries) \cite{Peskin:2001rw}. We then compare with the model contributions to $S$ and $T$. 
Fig.~\ref{st} shows various confidence regions for $S$ and $T$ (relative to a SM Higgs reference mass of 100 GeV), together with the model contributions in regions of parameter space with natural stop mass above 2 TeV (estimated from (\ref{stop}) with $D_{h,H,Z} \leq 4$, for the sake of argument), and satisfying the $\Gamma (Z \rightarrow b\bar{b})$ and $b \rightarrow s\gamma$ constraints. 
In Fig.~\ref{hb},
we plot the Higgs masses {\it versus} $\tan \beta$ for points with $m_{\tilde{t}}> 2$ TeV which are also compatible with electroweak constraints at the 95 $\%$ C.L.. Naturally large stop masses are obtained for $0.4 < \tan \beta < 2.5$, $2.9 < \lambda < 3.4$, $600 < m_H/\text{GeV} < 2800$ and $350 < m_h/\text{GeV} < 500$. As is clear from Fig. \ref{st}, large values of $m_h$ are compatible with electroweak precision tests for $\tan \beta$ not too close to unity, because there are large, positive contributions to $T$, coming from the other Higgs scalars. (For $\tan \beta =1$, custodial symmetry is restored and the contributions to $T$ cancel.)

Can an extended Higgs sector of this type be discovered at the LHC? Both neutral Higgs scalars lie above the threshold for the `gold-plated' decay $H,h \rightarrow ZZ \rightarrow l \bar{l}l \bar{l}$ but the width is shared with the decay to $t \bar{t}$, which has a large background \cite{gun3}. Thus, it is not clear that either $h$ or $H$ will necessarily be discovered in this scenario. 

In conclusion, we see that it is conceivable that supersymmetry solves the hierarchy problem, yet that only an extended Higgs sector will be accessible at the LHC. Furthermore, although we have presented just one model of this type, it is clear, from the introductory arguments, that many such models should exist. Thus, even though the experimental signatures of the specific model presented above should be reasonably straightforward to identify, it may be far more difficult in practice to determine whether or not supersymmetry solves the hierarchy problem if the LHC only discovers a Higgs sector.

Finally, let us re-consider the validity of the simplifying assumptions we have made. Firstly, we did not consider the full set of soft supersymmetry-breaking terms, in that we set $A=C=0$. Both of the effects described following Eq.\ref{stop}, \emph{viz.} the largeness of Higgs masses and the smallness of the logarithm in (\ref{stop}), are present in the region of non-vanishing $A$ or $C$, and we expect improved naturalness to persist as well. Secondly, we neglected corrections of $O(M^2_Z/m^2_h)$ to the Higgs spectrum, which correct the masses by a few {\it per cent}. A fuller analysis including these tree-level contributions, together with the full set of loop corrections, which also give corrections of a few {\it per cent} to the Higgs masses, and induce sensitivity to heavy thresholds other than $m_{\tilde{t}}$, would be welcome.

\begin{acknowledgments}
We thank J.~March-Russell, J.~F. Gunion, G.~G. Ross and T.~E.~J. Underwood for discussions.
\end{acknowledgments}


\begin{thebibliography}{10}
\bibitem{note4}
A truly realistic model with all Higgsino masses at the supersymmetry scale would require $A$ and $C$ to be non-zero.
\end{thebibliography}
\end{document}